\documentclass[letterpaper, 10 pt, conference]{ieeeconf}

\IEEEoverridecommandlockouts

\usepackage[T1]{fontenc}
\usepackage{cite}
\usepackage{amsmath,amssymb,amsfonts}
\usepackage{algorithmic}
\usepackage{graphicx}
\usepackage{textcomp}
\usepackage{xcolor}
\usepackage{subcaption}
\usepackage{ragged2e}
\usepackage{url}
\usepackage{setspace}
\usepackage{orcidlink}
\usepackage[capitalize]{cleveref}
\usepackage{tikz} 
\usepackage{eso-pic} 

\crefname{section}{Sec.}{Secs.}
\Crefname{section}{Section}{Sections}
\Crefname{figure}{Figure}{Figures}
\crefname{figure}{Fig.}{Figs.}
\Crefname{table}{Table}{Tables}
\crefname{table}{Tab.}{Tabs.}

\title{\LARGE \bf
Distance Estimation in Outdoor Driving Environments Using Phase-only Correlation Method with Event Cameras
}

\author{%
Masataka Kobayashi$^{1}$, Shintaro Shiba$^{2}$, Quan Kong$^{2}$, Norimasa Kobori$^{2}$,\\
Tsukasa Shimizu$^{3}$, Shan Lu$^{1}$, Takaya Yamazato$^{1}$\\
$^{1}$School of Engineering, Nagoya University, Nagoya, Japan\\
kobayashi@katayama.nuee.nagoya-u.ac.jp, \{lu, yamazato\}@yamazato.nuee.nagoya-u.ac.jp\\
$^{2}$Woven by Toyota, Inc., Tokyo, Japan\\
\{shintaro.shiba, quan.kong, norimasa.kobori\}@woven.toyota\\
$^{3}$TOYOTA MOTOR CORPORATION, Toyota, Japan\\
tsukasa\_shimizu@mail.toyota.co.jp
}

\newcommand{\AtPageUpperCenter}[1]{%
  \AddToShipoutPictureBG*{%
    \AtPageUpperLeft{%
      \raisebox{-1.5cm}{%
        \makebox[\paperwidth]{\begin{minipage}{\paperwidth}\centering #1\end{minipage}}}}}
}

\AtPageUpperCenter{%
    \begin{tikzpicture}[remember picture,overlay]
        \node[align=center,text=gray,font=\small] at (current page.north) [yshift=-1.5cm] {
            This paper has been accepted for publication at the IEEE Intelligent Vehicles Symposium (IV),\\
            Napoca, Romania, 2025. ©IEEE
        };
    \end{tikzpicture}
}
\begin{document}

\maketitle

\begin{abstract}

With the global proliferation of autonomous driving technology, the advancement of sensor technology is essential to ensure its safety. In particular, sensor fusion plays a crucial role in autonomous driving systems. However, equipping vehicles with multiple sensors leads to increased costs, making it necessary to have a single sensor that can perform multiple roles.

This study focuses on event cameras, exploring their potential among various sensor technologies. Event cameras possess characteristics such as high dynamic range, low latency, and high temporal resolution, and they can also leverage visible light communication. This enables high visibility in low-light and backlit environments, as well as excellent performance in detecting pedestrian movements and acquiring traffic information between traffic lights and vehicles. These characteristics are particularly beneficial for autonomous driving systems. Additionally, if distance estimation functionality can be provided by the event camera, it allows a single sensor to perform multiple roles, offering significant advantages in terms of cost efficiency.

In this study, we achieved distance estimation based on triangulation using an event camera and two points on an LED bar installed along a road. Furthermore, by employing the phase-only correlation method, we achieved sub-pixel precision in estimating the distance between two points on the LED bar, enabling even more accurate distance estimation. This approach performed monocular distance estimation in outdoor driving environments at distances ranging from 20 to 60 meters, achieving a success rate of over 90 \% with errors of less than 0.5 meters.

We are considering implementing position estimation in the future, with the current distance estimation technology forming the foundation for this. By achieving high-precision distance estimation, the vehicle's position relative to surrounding ITS smart poles can be accurately determined, enabling more precise position estimation. This will allow autonomous vehicles to know their exact position in real-time and select the optimal driving route based on surrounding traffic conditions and road conditions. Ultimately, we believe that this technology can contribute to the development of a smart transportation system in the city.

\end{abstract}

\section{INTRODUCTION}

Autonomous vehicle (AV) technology has rapidly grown over the past decade, and the expansion of autonomous driving has become a global topic. According to estimates from the U.S. National Highway Traffic Safety Administration (NHTSA), human error accounts for more than 90\% of traffic accidents, and with the spread of autonomous driving technology, a reduction in accidents caused by this factor is expected. There are already autonomous taxis in operation in certain areas without safety drivers and have achieved millions of miles with a lower accident rate compared to human-driven vehicles ~\cite{ArxivDec2024}.

One of the key components supporting autonomous driving technology is sensors. Autonomous robots must constantly monitor and avoid surrounding pedestrians, cyclists, and obstacles, where sensor capabilities play a vital role~\cite{ITS2024}. Current typical sensor sets for autonomous vehicles are GPS, inertial measurement units, RGB cameras,ultrasonic sensors, LiDAR, and radar~\cite{ICIEA2020}. Each of these sensors has its own strengths and weaknesses, and since no single sensor can meet all the stringent requirements for reliable operation, sensor fusion is necessary. However, increasing the number of sensors leads to higher costs, so the ideal solution is to achieve the roles of multiple sensors with a single sensor.

This study focuses on an event camera \cite{Lichtsteiner2006isscc,Gallego2022pami,Arxiv2024,Gehrig2024nature}, which is capable of receiving high spatio-temporal resolution while maintaining low data volume and is used as a visible light communication (VLC) signal receiver.  Event cameras possess rapid response characteristics, allowing them to react quickly to sudden movements, such as obstacles and pedestrians darting into traffic ~\cite{Arxiv1906}. Moreover, with a high dynamic range of over 120 dB, they perform well even under challenging lighting conditions such as nighttime or backlight. Event cameras are also compatible with VLC, which opens up new possibilities in autonomous driving~\cite{Katayama2022}~\cite{Arxiv202}~\cite{VNC2018}~\cite{Katayama2691}. In VLC, information is transmitted by the high-frequency blinking of LEDs, and the event camera detects it with high temporal resolution. This allows vehicles to receive real-time information from smart poles and traffic lights on the road and incorporate this data into driving control, enabling safe and efficient driving.

The objective of this study is to add distance estimation capabilities to the event camera with the aforementioned features. Previous studies on obstacle avoidance have primarily focused on reducing object detection time\cite{ICICC111}, object recognition, and motion prediction~\cite{Arxiv1906}, with less emphasis on distance estimation to obstacles. However, in driving environments, it is crucial not only to recognize and avoid obstacles but also to obtain accurate distance information, as this is essential for enhancing the safety and efficiency of autonomous systems. By incorporating distance estimation, improvements in collision avoidance, maintaining safe following distances, and enhancing the accuracy of emergency braking and steering operations can be expected. Furthermore, distance estimation using event cameras is faster than LiDAR, enabling quicker responses in dynamic driving environments. The system will also demonstrate stable performance in driving environments, long-range scenarios, and situations where GPS is unavailable. By integrating multiple functions into a single sensor, the system's cost can be reduced as well. This technology holds the potential to significantly enhance the safety and efficiency of autonomous vehicles.

This study aims to estimate distances of over 20 meters in outdoor driving environments. This is in contrast to previous research on distance estimation using event cameras, which has predominantly focused on indoor environments~\cite{arxiv2107}~\cite{Katayama2016} and short-range estimation~\cite{arxiv2107}.
Specifically, a distance estimation system using triangulation was developed with two points on an LED bar light installed on the road. To perform distance estimation using triangulation, it is necessary to calculate the pixel distance between two points on the LED bar in the event data. Since this pixel distance directly affects the estimation accuracy, a POC was applied to measure the pixel distance down to the subpixel range.

In outdoor mobile experiments, a vehicle equipped with a camera moved at a speed of 5.2 m/s, and distance estimation was performed over a range of 20 to 60 meters. The results showed that estimation with an error of less than 0.5 meters was achieved in over 90\% of the locations. These results demonstrate the effectiveness of the triangulation-based distance estimation method and the feasibility of achieving high-precision distance measurement using event cameras.

This study makes the following contributions:
\begin{itemize}
    \item  Based on triangulation, the phase-only correlation (POC) method is applied to the separated event data to estimate the distance between LEDs at a sub-pixel level.
    \item  In outdoor driving experiments, a measurement error of less than 0.5 meters was achieved in 90\% of cases at distances between 20 meters and 60 meters at a speed of 20 km/h.
    
    \item  At a speed of 30 km/h, a measurement error of less than 0.5 meters was achieved in 83.7\% of cases at distances between 20 meters and 55 meters.
\end{itemize}
\section{System Model}
This section explains the system model used in this study. The distance estimation is performed using an event camera as the receiver and an LED bar light as the transmitter. The event camera used is the silkyEvCam equipped with the Sony IMX636 sensor (resolution: 1280 $\times$ 720
, temporal resolution: below 100 $\mu$ s), and the software utilized is the METAVISION SDK. The LED bar light consists of 96 red LEDs arranged at 1 cm intervals, with each LED's blinking frequency being independently adjustable.Increasing the interval between LEDs from 1 cm to a longer distance, thereby extending the baseline, can theoretically improve the accuracy of triangulation. In triangulation-based distance estimation, the physical distance between two points (the baseline) and the pixel distance between their corresponding image points are used to calculate the target distance. As the baseline increases, the effect of pixel-level errors on the estimated distance decreases, leading to potentially higher estimation accuracy.
However, a longer LED spacing results in a longer LED bar, which may exceed the camera's field of view and require additional adjustments such as lens focal length or installation position. Therefore, a trade-off must be considered between improving accuracy through baseline extension and maintaining ease of implementation. The LED bar light is used in a vertical orientation, with a total of 10 LEDs used from the top and bottom sections. In the future, it is expected that each LED will transmit different information in the context of visible light communication. Therefore, a system capable of simultaneous visible light communication and distance measurement is assumed, employing different blinking frequencies for the LEDs. The blinking frequencies for both the upper and lower sections are 5000 Hz, 10000 Hz, 20000 Hz, 10000 Hz, and 5000 Hz.

\begin{figure}[t]
    \centering
    \includegraphics[width = \linewidth]{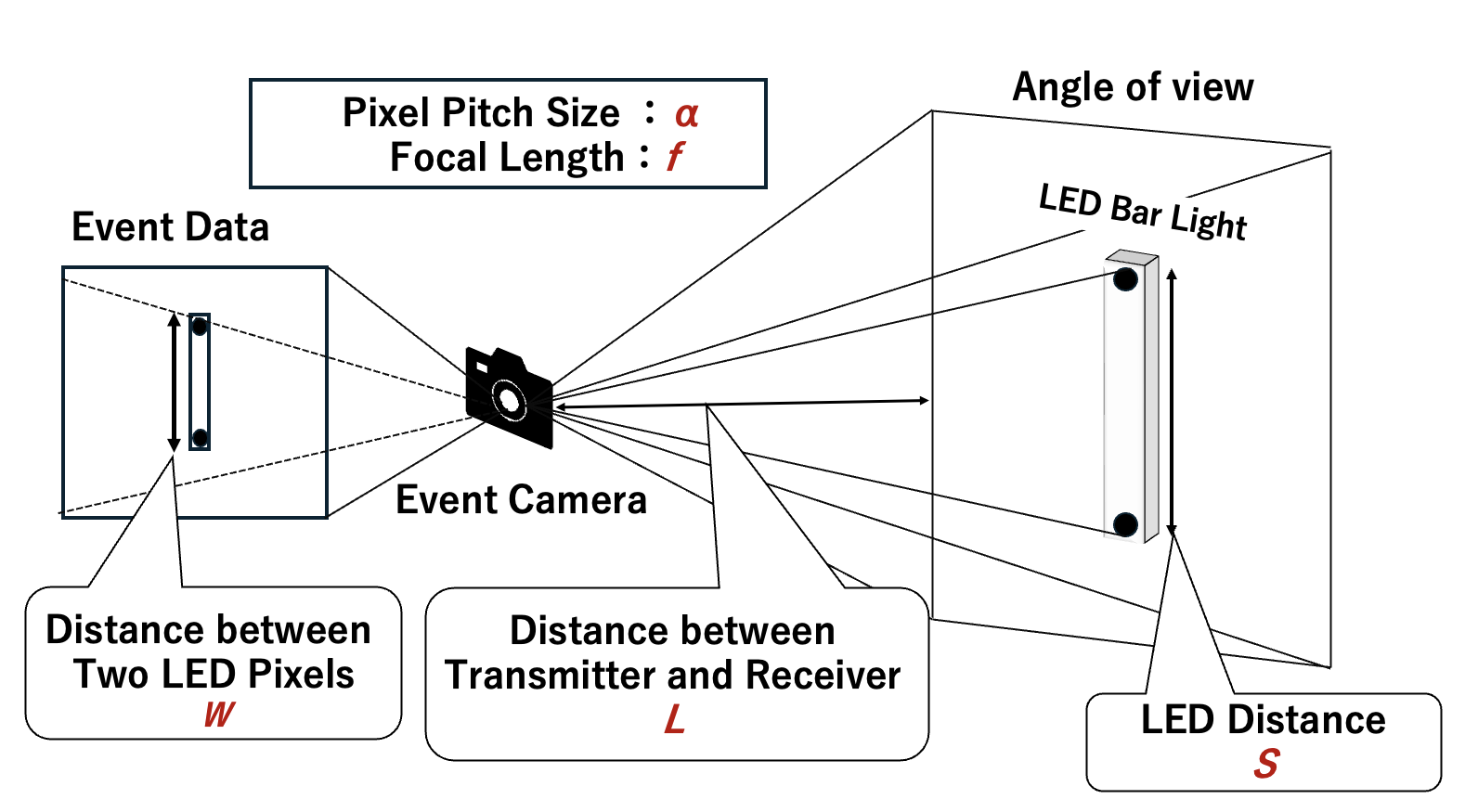}
    \vspace{-4mm}
    \caption{Triangulation model}
    \label{fig:tri}
\end{figure}

The distance estimation method in this study is based on triangulation. To estimate the distance between the transmitter and receiver, four parameters are used: pixel pitch size $\alpha$, focal length $f$, the actual distance between the two LEDs $S$, and the pixel distance $W$ between the two LEDs. 
(Fig.~\ref{fig:tri})
\begin{equation}
 L= \frac{fS}{W\alpha}
\end{equation}
The distance between the top and bottom LEDs is used as the inter-LED distance to reduce the absolute error in distance estimation caused by errors in $W$, which is important for triangulation-based distance estimation. Since the three parameters other than $W$ are constant, the distance L between the transmitter and receiver depends solely on $W$. Therefore, the primary goal of this study is to determine the pixel distance $W$ from the event data. However, since the value of W derived from this calculation is in pixel units and can only hold integer values, subpixel precision, or estimation of distances smaller than 1 pixel, is required. In this study, phase correlation is used to achieve subpixel-level accuracy in the estimation.

\begin{figure*}[tbp]
\centering
\includegraphics[width=\linewidth]{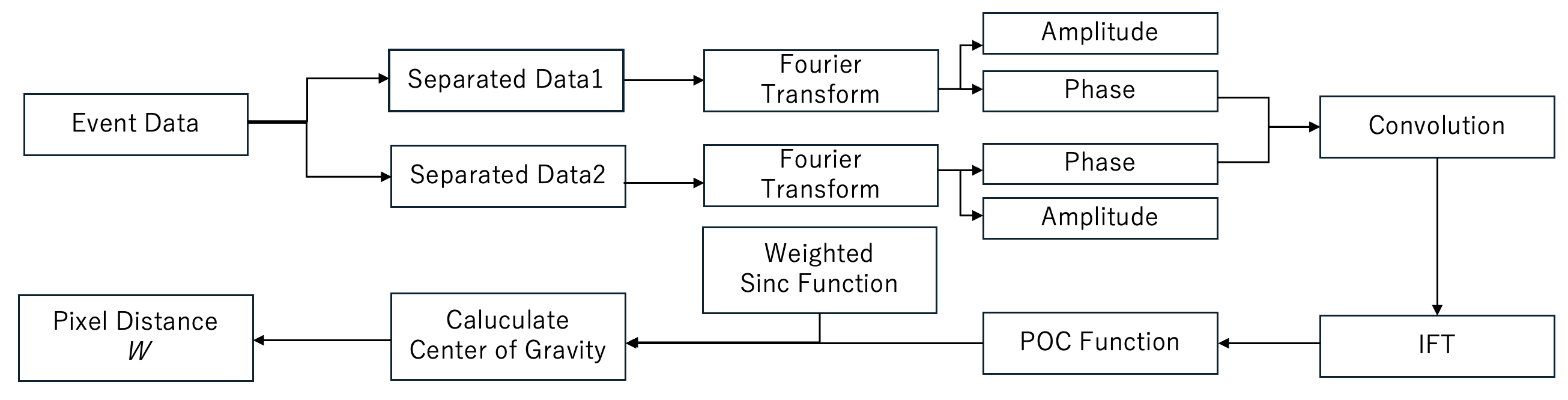}
\caption{System model
}
\label{fig:systemmodel}
\end{figure*}

\subsection{Processing and Separation of Event Data}

The phase correlation method is a technique used to accurately determine the displacement between two images based on their intensity values. However, in the case of event cameras, only events triggered by changes in intensity are obtained, rather than intensity values themselves. Therefore, instead of using intensity values, the number of events occurring within a specified time window at each pixel is used to generate a 2D array. By using the event count, the phase correlation method can be applied in a similar manner to traditional phase correlation techniques.
Specifically, the number of events occurring at all pixels within a specified time window is measured, and a 2D array of event counts is created. Images are typically represented as 2D arrays of intensity values, and by replacing them with an array of event counts, the phase correlation method can be applied. Just as the intensity from an LED decreases as it moves away from the center, the number of events also decreases as the distance from the center increases, making it less likely to exceed the threshold, and the number of events diminishes. As a result, there is a strong compatibility between the intensity-based correlation method and the characteristics of the event camera, making it suitable for applying correlation techniques.

Furthermore, to apply the phase correlation method, it is necessary to separate the event data, which records the blinking of the upper and lower 5 LEDs of the LED bar, into the upper and lower sections. In this study, we utilize the property that the weighted average coordinate of the LED's y-axis always falls between two LEDs. This coordinate is used as the boundary to separate the event data. 
\begin{equation}
\bar{y}_{\text{weighted}} = \frac{\sum_{i=1}^{N} y_i \cdot w_i}{\sum_{i=1}^{N} w_i}
\end{equation}
where $\bar{y}_{\text{weighted}}$ represents the weighted average y-coordinate, $N$ is the total number of events (data points), $y_i$ is the y-coordinate of the $i$-th event, and $w_i$ is the weight of the $i$-th event (the number of events recorded at that pixel).
This approach eliminates the need for setting an appropriate threshold, unlike methods that filter pixels with high event counts based on a threshold, and allows reliable separation of the two points regardless of the distance.

\subsection{POC processing}

\begin{figure}[htbp]
    \centering
    \includegraphics[height=5cm, keepaspectratio]{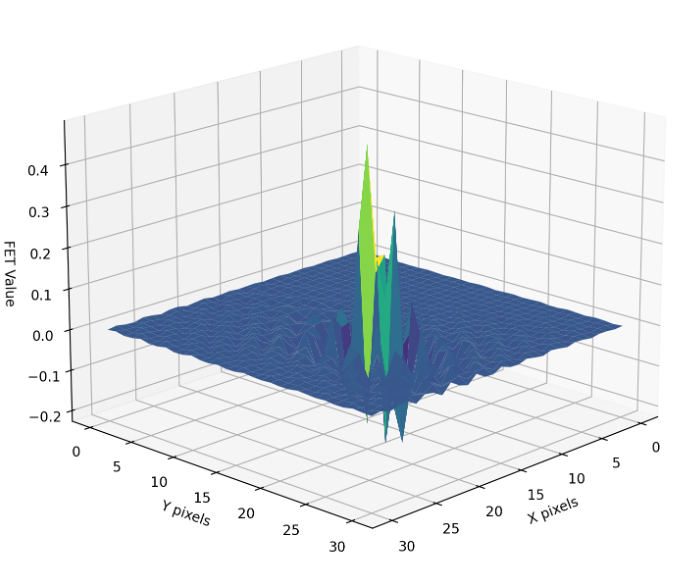} 
   
    \caption{POCFunc(\( G_{12} \)) }
    \vspace{-4mm}
    \label{fig:poc}
\end{figure}

The following section explains how to apply the phase-only correlation method to two separate two-dimensional event count arrays. Each two-dimensional array is assumed to be represented as an \( M \times N \) matrix, denoted as \( f_1(m,n) \) and \( f_2(m,n) \), respectively.
The results of applying the discrete Fourier transform to them are defined as \(F_1(u,v)\)and \( F_2(u,v)\).
\begin{equation}
F_1(u,v) = \sum_{m=0}^{M-1} \sum_{n=0}^{N-1} f_1(m,n) e^{-j 2\pi \left( \frac{mu}{M} + \frac{nv}{N} \right)}
\end{equation}

\begin{equation}
F_2(u,v) = \sum_{m=0}^{M-1} \sum_{n=0}^{N-1} f_2(m,n) e^{-j 2\pi \left( \frac{mu}{M} + \frac{nv}{N} \right)} 
\end{equation}
This method allows for the acquisition of spatial frequency spectra for each two-dimensional array. After computing the convolution of the obtained \( F_1 \) and \( F_2 \), dividing by the absolute values of their amplitudes yields \( J_{ab} \).
\begin{equation}
{J}_{\text{12}} = \frac{F_1(u,v)F_2^*(u,v)}{\left| F_1(u,v)F_2^*(u,v) \right|
}
\end{equation}
Here, one of the terms in the multiplication is taken as the complex conjugate.By applying the inverse discrete Fourier transform to \( J \), the phase-only correlation function \( G_{12} \) (Fig.~\ref{fig:poc})is obtained.

\begin{equation}
G_{12}(m,n) = \frac{1}{MN} \sum_{u=0}^{M-1} \sum_{v=0}^{N-1} J_{12}(u,v) e^{j 2\pi \left( \frac{mu}{M} + \frac{nv}{N} \right)}
\end{equation}
\begin{figure}[htbp]
    \centering
    \includegraphics[height=5cm, keepaspectratio]{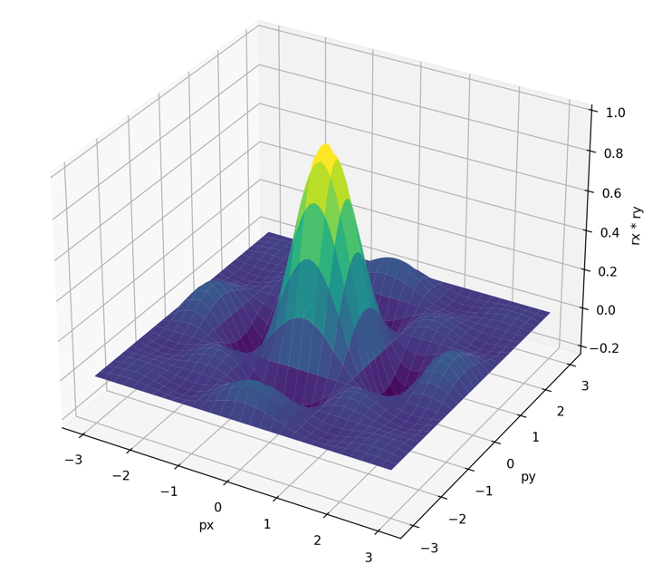} 
    
    \caption{sincFunc}
    \vspace{-4mm}
    \label{fig:sinc}
\end{figure}

The \(x\) and \(y\) coordinates where the phase-only correlation (POC) function exhibits the highest value represent the integer-valued pixel distance \(W\) within the event data. A weighting method utilizing the sinc function is applied to compute subpixel-level precision.

From the POC function, a \(5 \times 5\) region $N$centered around the peak position is extracted. 
\begin{equation}
N = C[y_p - 2 : y_p + 2, \, x_p - 2 : x_p + 2]
\end{equation}
For each pixel value within this region, 1D sinc function $w_x$ and $w_y$   are applied to calculate the weights. 

The sinc function is designed to assign higher weights to values near the center, with weights decreasing as the distance increases. The sinc function(Fig.~\ref{fig:sinc}) is applied in both the horizontal and vertical directions and the resulting weights are combined to create a 2D weighting matrix$W$.
\begin{equation}
w_x = \text{sinc}(x), \quad w_y = \text{sinc}(y)
\end{equation}
\vspace{-4mm}
\begin{equation}
W = w_x \otimes w_y
\end{equation}

Using this weighting matrix, the correlation values within the neighborhood are adjusted, and the center of gravity is calculated in both the \(x\) and \(y\) directions, resulting in the determination of \(x_{\text{peak}}\) and \(y_{\text{peak}}\).

\begin{equation}
x_{\text{peak}} = \frac{\sum_{i,j} N[i,j] \cdot W[i,j]}{\sum_{i,j} N[i,j] \cdot W[i,j] \cdot x_j}
\end{equation}
\begin{equation}
 y_{\text{peak}} = \frac{\sum_{i,j} N[i,j] \cdot W[i,j]}{\sum_{i,j} N[i,j] \cdot W[i,j] \cdot y_j}
\end{equation}

This approach allows for subpixel precision in determining the peak position. It achieves a level of accuracy that cannot be attained through simple maximum value detection alone.

\section{Considerations in a Moving Environment}

In this experiment, distance estimation in outdoor moving environments was conducted by applying the system model mentioned above. There are several factors to consider when conducting experiments in moving environments. The first factor is the influence of vibration. The vibration received by the event camera mounted on the vehicle is primarily caused by uneven road surfaces. According to prior research, when a high-speed camera with a 35 mm focal length lens and a resolution of 512$\times$512 pixels is mounted on a vehicle traveling at 30 km/h, vertical movement is likely to occur, with up to 1.5 pixels of movement per millisecond.\cite{Katayama201}. Additionally, the camera is also somewhat affected by engine vibrations. To reduce the influence of engine vibrations, the camera was mounted on the side of the vehicle rather than on the bonnet, which increased the distance from the engine. Although the phase-only correlation method can detect movement and vibration that occur within a 0.01-second period, it is difficult to improve the accuracy of the method itself, as it cannot directly address the number of events occurring. Therefore, it is considered difficult to improve the accuracy of the phase-only correlation method. The key direction for improving accuracy is to minimize the accumulation time. In this experiment, the accumulation time was set to 3000 $\mu$s for data analysis.

The second factor is background noise caused by the movement of the camera itself. The number of background noise events in a vehicle-mounted driving environment is enormous, and when it exceeds 1.06 GEPS (Giga event per second), data loss occurs. Therefore, in this study, to reduce the number of events, a high-pass filter\cite{arxiv2210}  from the METAVISION SDK was used. Even after applying the high-pass filter, events that remained were further filtered by an event-count filter. Events that did not exceed the threshold were set to zero and removed.

\section{Distance Estimation Experiment in Outdoor Dynamic Environment} 
\begin{figure}[t]
    \centering
    \includegraphics[width = \linewidth]{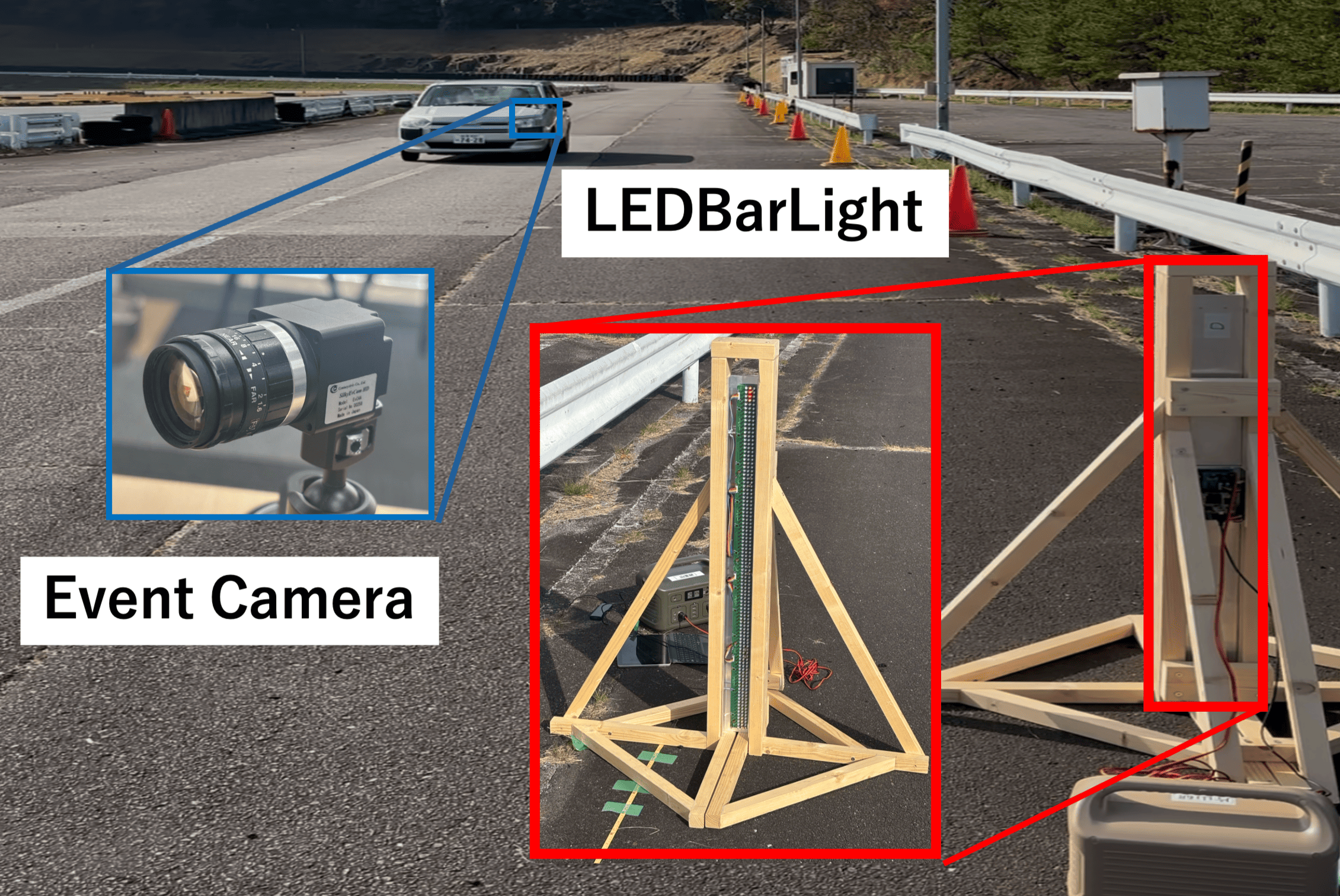}
    \vspace{-4mm}
    \caption{Experimental environment}
    \label{fig:ev}
\end{figure}

\subsection{True value measurement method}
In a stationary environment, the true distance from the transmitter can always be measured with a tape measure. However, in a moving environment, it is necessary to develop a method to accurately measure the true distance. In previous studies\cite{Katayama2849}, pylons were placed at regular intervals from the transmitter, and the locations where the pylons would appear at the edge of the camera’s field of view at specific distances were determined in advance. By doing so, the true distance could be known during image analysis. However, this method is highly susceptible to errors caused by lateral shifts in the vehicle’s position, which significantly affect the accuracy of the true distance measurement.

To address this issue, this study used a synchronized smartphone camera with an event camera to constantly capture a pylon indicating the distance from the transmitter. By recording the elapsed time since the start of the capture and the vehicle’s position, the true distance was measured. To synchronize the two cameras, an LED array with 16 LEDs arranged in a square (one side containing 16 LEDs) was used. This LED array alternated between 5 seconds of being off and 1 second of blinking. The event camera mounted on the vehicle’s side and the smartphone camera fixed on the passenger seat captured the same blinking cycle, achieving synchronization.

A smartphone camera with 60 fps was used instead of a high-speed camera with a high temporal resolution because running two cameras simultaneously on a single computer posed challenges in terms of the vast amount of data and processing load, which could result in frame loss and data transmission delays. Additionally, since the pylons were placed only 2.5 meters away from the vehicle, the required resolution was not high. However, due to the 60 fps limitation, there is a potential error of up to 8 cm at a speed of 20 km/h, so the use of a smartphone camera with a higher frame rate is being considered.

\subsection{Experiment Results}

\begin{figure}[htbp]
    \centering
    \includegraphics[width = \linewidth]{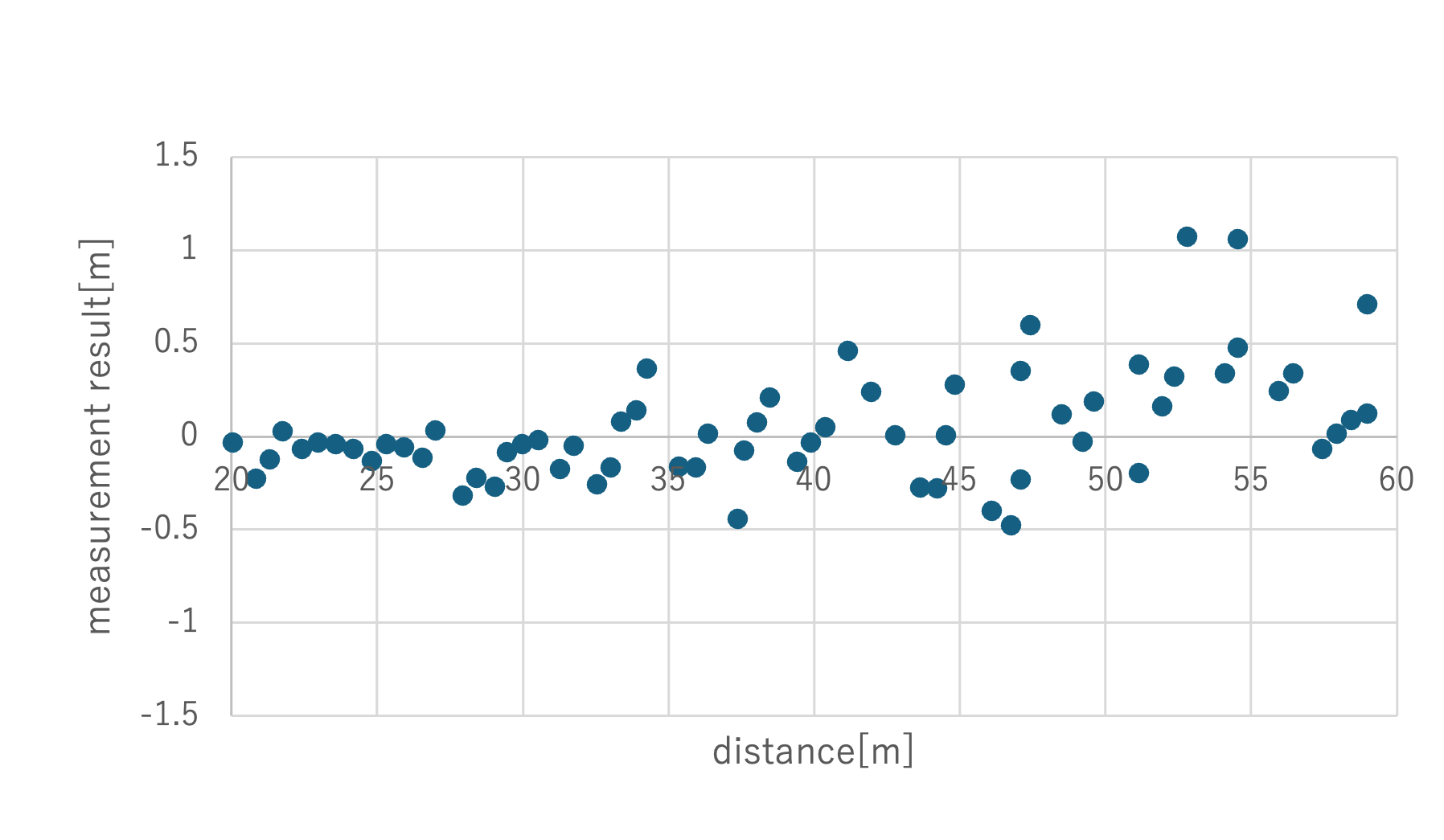}
    \vspace{-4mm}
    \caption{Estimated distance and measurement error at 20 km/h}
    \label{fig:20km}
\end{figure}

\begin{figure}[htbp]
    \centering
    \includegraphics[width = \linewidth]{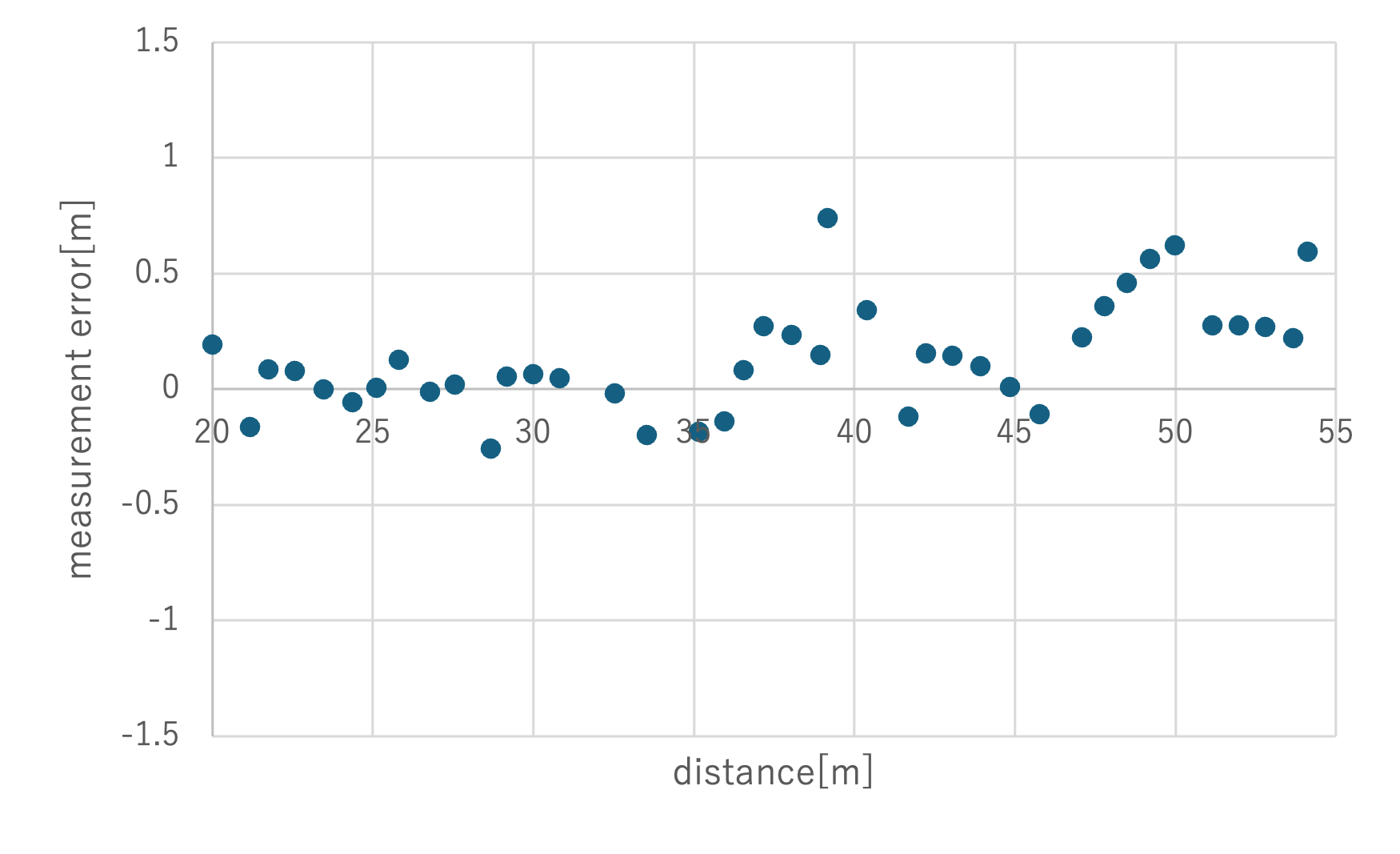}
    \vspace{-4mm}
    \caption{Estimated distance and measurement error at 30 km/h}
    \label{fig:30km}
\end{figure}

TBased on the above considerations, an outdoor mobile experiment was conducted. The experiment took place at Kyosei Traffic University in Okazaki City, Aichi Prefecture. The high-speed outercourse used for the experiment had a surface with a significant amount of gravel, and it is expected that the vibration effects were more pronounced compared to public roads. The lens used had a focal length of 35 mm. The distance estimation was conducted at speeds of 20 km/h for distances ranging from 20 m to 60 m and at 30 km/h for distances ranging from 20 m to 55 m. The following shows pictures of the experimental scene, the event camera, and the LED bar light.(Fig.~\ref{fig:ev}) 
As a result, at a speed of 20 km/h, a measurement error of 0.5 m or less was achieved in 90\% of the distances between 20 m and 60 m from the transmitter. (Fig.~\ref{fig:20km})  Similarly, at a speed of 30 km/h, an error of 0.5 m or less was achieved in 83.7\% of the distances between 20 m and 55 m. (Fig.~\ref{fig:30km})  These results demonstrate that high-accuracy distance estimation is achievable at distances of 20 m or more, even in outdoor mobile environments.


\section{Conclusion}
\label{sec:conclution}
This study proposes a highly accurate distance estimation method using an event camera and demonstrates its reliable performance even in outdoor driving environments. The proposed method is based on triangulation and applies the phase-only correlation (POC) method to the separated event data, achieving sub-pixel level measurement of the distance between LEDs. In outdoor mobile experiments, the method achieved a measurement error of 0.5 m or less in 90\% of cases at distances between 20 m and 60 m at a speed of 20 km/h and in 83.7\% of cases at distances between 20 m and 55 m at a speed of 30 km/h. Notably, the proposed method maintains stable and high-accuracy distance estimation even under dynamic outdoor conditions. This capability enables practical distance estimation in environments that were challenging for conventional technologies, opening new possibilities for applications such as autonomous driving and robotic vision.

The proposed method is expected to be applied to cooperative control for intersection entry at traffic signals and complementary positioning in environments where GPS is unstable, such as tunnels or urban areas.Future challenges include further improving accuracy (to achieve a measurement error of 0.1 m or less), extending the measurable distance range (up to 100 m), and enhancing tolerance to high-speed movement in dynamic environments (exceeding 45 km/h). Addressing these challenges will further expand the potential of sensor technologies utilizing event cameras and significantly enhance their practical applications in fields such as autonomous driving and robotics. The outcomes of this study represent a critical step toward groundbreaking innovations, paving the way for future advancements in these domains.

\addtolength{\textheight}{-12cm}   

\section*{Acknoledgement}
We would like to thank H. Okada (Nagoya Univ.) for their useful discussions.

\end{document}